\documentstyle[12pt]{article}

\newcommand{\semi}{;\hfil\break}

\newcommand{\eq}{\begin{equation}}
\newcommand{\eqe}{\end{equation}}
\newcommand{\eqa}{\begin{eqnarray}}
\newcommand{\eqae}{\end{eqnarray}}

\newcommand{\del}{\partial}

\begin{document}

\pagestyle{empty}

\hfill{NSF-ITP-94-53}

\hfill{hep-th/9408005}

\hfill{July 1994}

\vspace{18pt}

\begin{center}
{\large \bf CRITICAL BEHAVIOR OF THE MARINARI-PARISI MODEL}
\vspace{16pt}

{\bf Shyamoli Chaudhuri and Joseph Polchinski}

\vspace{16pt}
\sl

Institute for Theoretical Physics \\ University of California \\
Santa Barbara, California 93106-4030

\rm

\vspace{12pt}
{\bf ABSTRACT}

\end{center}

\begin{minipage}{4.8in}
We consider the continuum string theory corresponding to the Marinari-Parisi
supersymmetric matrix model.  We argue that the world-sheet physics is exotic,
and different from any known supersymmetric string theory.  The embedding
superspace coordinates become disordered on the world-sheet, but because
of the noncompactness of the embedding time the disorder
becomes complete only at asymptotic world-sheet scales.

\end{minipage}

\vfill

\pagebreak
\pagestyle{plain}
\setcounter{page}{1}
\setcounter{section}{-1}
\baselineskip=16pt

Understanding the nature of string theory beyond perturbation theory remains
one of the key problems in fundamental theory.
Matrix models are one of the few means of obtaining insight into
this question,\footnote
{For reviews, see ref.~\cite{mmreviews}.}
having revealed, for example, the existence of unexpectedly large
nonperturbative string effects\cite{shenker}.    Many important
non-perturbative
issues, especially the structure of the vacuum, are likely to involve spacetime
supersymmetry in an essential way.  A simple spacetime supersymmetric matrix
model has been formulated by Marinari and Parisi\cite{marpar},
but there has been relatively little further
development\cite{dhab,otros,coll,collj}.  In particular, the continuum string
theory which corresponds to the scaling limit of the Marinari-Parisi model has
never been identified.  In this paper we partially
resolve this issue.  The result is surprising, with rather exotic world-sheet
physics, and it is different from any known supersymmetric string theory.

The Marinari-Parisi model\cite{marpar} is the quantum mechanics of an $N\times
N$ matrix in a one-dimensional superspace
\begin{equation}
{\bf\Phi}(\tau,\theta,\bar\theta) = {\bf M}(\tau) + \bar\theta{\bf \Psi}(\tau)
+ {\bf \bar\Psi}(\tau)\theta + \theta\bar\theta {\bf F}(\tau).
\end{equation}
The action is\footnote
{We take the embedding time $\tau$
to be Euclidean, so that the world-sheet path integral is a convergent
gaussian;
Minkowski amplitudes are as usual defined by continuation.}
\begin{equation}
S = -N \int d\tau\,d\bar\theta\,d\theta\, {\rm Tr}\biggl\{
\frac{1}{2} \bar D {\bf\Phi} D {\bf\Phi} + W({\bf\Phi}) \biggr\}, \label{act}
\end{equation}
with $D = \del_{\bar\theta} + \theta \del_\tau$,
$\bar D = -\del_{\theta} - \bar\theta \del_\tau$.
The Feynman graph expansion for this model generates a discretization of random
surfaces in superspace.  Marinari and Parisi\cite{marpar} solved the
zero fermion number sector by diagonalizing
${\bf M}$ and rewriting the theory in terms
of non-relativistic free fermions \cite{BIPZ}.
This was done for the cubic superpotential $W({\bf\Phi})
= \frac{1}{2} {\bf\Phi}^2 - \frac{1}{3} \lambda {\bf\Phi}^3$, which is
much simpler than the general polynomial case, and has been shown to
have the generic critical behavior of this class of potentials \cite{collj}.
Dabholkar\cite{dhab} extended this to include those
components of ${\bf \Psi}$ and ${\bf \bar\Psi}$ which are diagonal in the same
basis as ${\bf M}$, and obtained a supersymmetric extension of the free
fermi theory.

The model with cubic superpotential has a critical point at $\lambda_c^2
= 1/6 \sqrt{3}$, where the perturbation series diverges and large
graphs dominate.  In the double-scaling limit, $\lambda \to \lambda_c$
with the renormalized inverse string coupling,
$\kappa^{-1}$ $=$ $N(\lambda_c - \lambda)^{5/2}$
held fixed, graphs of all topologies survive.  The critical behavior
is then expected to be described by some spacetime supersymmetric string
theory.  Various
possibilities, and difficulties associated with each, are discussed in
ref.~\cite{dhab}.  Rather than repeat these arguments here, we will point out a
puzzle that has not previously been emphasized.

In order to obtain a nontrivial double-scaling limit, it is
necessary\cite{dhab} to rescale the
embedding time, holding
\begin{equation}
\tau(\lambda_c - \lambda)^{1/4} \ =\ {\rm fixed}. \label{tauscale}
\end{equation}
This is
contrary to the situation in the bosonic $d=1$ theory\cite{c1refs}, where the
embedding time is not rescaled.  Moreover, it would appear to violate general
principles of quantum field theory: time translation is a global symmetry of
the
world-sheet theory, and so should not be renormalized\cite{nonren}.

To see what is happening, let us examine the expected continuum world-sheet
theory.  The superspace propagator for the theory~(\ref{act}) is
\begin{equation}
\frac{1}{2} ( \bar D D - D \bar D + 2 )(1 - \del_\tau^2)^{-1}.
\end{equation}
As usual in the matrix model, one expects the replacement of
$\exp(\del^2_\tau)$
for the propagator $(1 - \del_\tau^2)^{-1}$ to leave the theory in the
same universality class, since their infra-red behavior coincides.
The propagator between vertices at
superspace positions $(\tau,\theta,\bar\theta)$
and $(\tau',\theta',\bar\theta')$ becomes
\begin{equation}
\exp\left\{ -\frac{1}{2} \Bigl[ \tau'- \tau -
(\bar\theta' - \bar\theta) \theta +
\bar\theta (\theta' - \theta) \Bigr]^2 + (\theta' - \theta)
(\bar\theta' - \bar\theta) \right\}.
\end{equation}
The naive continuum limit gives the Euclidean action
\begin{equation}
S = \frac{z_1}{2}\Bigl(\del_a \tau - \del_a\bar\theta\theta + \bar\theta
\del_a\theta\Bigr)^2 - z_2 \del_a\theta\del_a\bar\theta, \label{contact}
\end{equation}
where we allow for general normalizations $z_1$ and $z_2$. This could also have
been written down directly as the most relevant action allowed by the
supersymmetry
\begin{equation}
\delta\theta = \epsilon, \qquad
\delta\bar\theta = \bar\epsilon, \qquad
\delta \tau = - \epsilon\bar\theta - \bar\epsilon\theta.
\end{equation}

The action~(\ref{contact}) is not free, and in fact is not conformally
invariant.  The one-loop beta functions can be computed as
\begin{equation}
\mu \del_\mu z_1 = g^2 z_1/\pi,\qquad
\mu \del_\mu z_2 = g^2 z_2/\pi, \label{beta}
\end{equation}
where the effective coupling $g^2 = z_1/z_2^2$ satisfies
$\mu \del_\mu g^2 = -2 g^4 /\pi$.  At longer world-sheet
distances, the coupling grows and both $\tau$ and $\theta$, $\bar\theta$
become more disordered.  There are two logical possibilities for the infrared
limit: a nontrivial fixed point or a mass gap.  In fact, we can largely
exclude the former possibility.  At a fixed point, one expects that
the spacetime translation and supersymmetries, being global symmetries of
the world-sheet theory, give rise to a current algebra,
\begin{eqnarray}
j_\theta(z) j_{\bar\theta}(0) &\sim& \frac{k}{z^2} + \frac{j_\tau(0)}{z}
\label{ope1}\\
j_\tau(z) j_\theta(0) &\sim& j_\tau(z) j_{\bar\theta}(0) \ \sim\ {\rm analytic}
\label{ope2}. \end{eqnarray}
However, taking the operator product of $j_\tau$ with~(\ref{ope1}) and
using~(\ref{ope2})
implies that the $j_\tau j_\tau$ operator product is analytic.  In
 a unitary theory,
this would strictly imply that $j_\tau$ is trivial.
In the present case the second
order fermionic action is not unitary, but the argument still suggests
that $j_\tau$ is trivial---that is, that there is a mass gap for world-sheet
states with nonzero spacetime frequency.  Since both beta
functions~(\ref{beta}) have the same sign, we expect that $\theta$ and
$\bar\theta$ also develop a mass gap.

The rather surprising conclusion is that the spacetime supersymmetry of the
matrix quantum mechanics problem does not survive in the critical
theory: the superspace coordinates become massive, leaving only the world-sheet
metric.  This accounts for a key feature of the exact solution.
The equal-time expectation values of the
Marinari-Parisi model are the same as the expectation values
in the $d=0$
bosonic string theory\cite{marpar}, both being given by
\begin{equation}
\int d{\bf \Phi}\, e^{- {\rm Tr}\{ W({\bf \Phi}) \} }\, \ldots\ .
\end{equation}
This equivalence now has the simple world-sheet explanation that
these theories are actually the same at long world-sheet distances.

This cannot, however, be the end of the story.  By the
rescaling~(\ref{tauscale})
of $\tau$, Dabholkar\cite{dhab} obtained a theory with non-trivial
spacetime dynamics, including nonperturbative supersymmetry breaking.
The rescaling, toward longer embedding-time scales as the critical point is
approached, agrees in direction with our argument that $\tau$ is becoming
disordered.  But how can a non-trivial dynamics be consistent with
a mass gap, since one would then expect all $\tau$-dependence to
disappear from the critical theory?  The key here is the unusual feature that
the $\tau$-coordinate is non-compact.  This is the only example of which we are
aware in which a non-compact field disorders.  For a compact field, there is
some length scale at which the disorder becomes comparable to the range of the
field and so one can say the field is completely disordered.  For a non-compact
field, however, one would expect that as one integrates to longer distances,
the fluctations grow, but that they do not become
infinite at any finite scale.  Thus, by scaling to longer embedding time scales
as the world-sheet cosmological canstant is taken to the critical point,
one can obtain a non-trivial limit.
In other words, for expectation values at fixed embedding time or frequency
scales there is a world-sheet mass gap, but the gap goes to zero as
the embedding time scale goes to zero.

Unfortunately we have not been able to make this picture more quantitative,
and have not been able to derive the scaling
exponent~(\ref{tauscale}).
The non-compactness of
$\tau$ makes the theory (\ref{contact}) appear complicated
even on a flat world-sheet, and we do not know of an analytic
approach.\footnote
{Various heuristic arguments have all led us to the incorrect scaling
$\tau \ \propto\ (\lambda_c - \lambda)^{-1/2}$.}
But we believe that our general picture---that
the superspace coordinates become disordered
at long world-sheet distances, but the embedding time becomes fully disordered
only asymptotically---is correct, both because of its
plausibility, and because of the difficulty of understanding the
rescaling~(\ref{tauscale})
of $\tau$ in any other way.

In conclusion, the Marinari-Parisi model has exotic world-sheet and
presumably also spacetime physics, but it is rather different from the
usual spacetime supersymmetric string theories. Most of these involve
world-sheet chirality.  It is possible to introduce chirality into the
matrix model vertices, and Siegel\cite{siegel} has proposed a discretization
of the Green-Schwarz superstring, but we are not aware of a solvable
example.

\centerline{\bf Acknowledgements}

This work was supported in part by
National Science Foundation grants PHY89-04035 and PHY91-16964.

\vfill

\pagebreak

\end{document}